\documentclass[twocolumn]{svjour3}         
\smartqed  
\usepackage{graphicx}
\usepackage{fix-cm}

\usepackage{times}
\usepackage{amsmath}
\usepackage{latexsym}

\usepackage{graphicx}
\usepackage{color}
\usepackage{dcolumn}
\usepackage{bm}
\usepackage{graphics}
\usepackage{subfigure}
\usepackage{timestamp}
\usepackage{multirow}

\usepackage{hyperref}

\def\comment#1{}

\journalname{Networking Science}

\begin{document}


\title{
Path Selection for Quantum Repeater Networks
}
\author{Rodney Van Meter
\and
Takahiko Satoh
\and
Thaddeus D. Ladd
\and\\
William J. Munro
\and
Kae Nemoto
}%
\institute{R. Van Meter \at
  Keio University, Fujisawa, Japan
  \email{rdv@sfc.wide.ad.jp}
  \and
  T. Satoh \at
  University of Tokyo, Japan
  \email{satoh@is.s.u-tokyo.ac.jp}
  \and
  T. D. Ladd \at
  Stanford University, Palo Alto, CA, and National Institute of
  Informatics, Tokyo, Japan
  \and
  W. J. Munro \at
  NTT Basic Research Labs, Atsugi, Japan
  \and
  K. Nemoto \at
  National Institute of Informatics, Tokyo, Japan
}

\date{Received: date / Accepted: date}

\maketitle

\begin{abstract}
  Quantum networks will support long-distance quantum key distribution
  (QKD) and distributed quantum computation, and are an active area of
  both experimental and theoretical research.  Here, we present an
  analysis of topologically complex networks of \emph{quantum
    repeaters} composed of heterogeneous links.  Quantum networks have
  fundamental behavioral differences from classical networks; the
  delicacy of quantum states makes a practical path selection
  algorithm imperative, but classical notions of resource utilization
  are not directly applicable, rendering known path selection
  mechanisms inadequate.  To adapt Dijkstra's algorithm for
  quantum repeater networks that generate entangled Bell pairs, we
  quantify the key differences and define a link cost metric, {\em
    seconds per Bell pair} of a particular fidelity, where a single
  Bell pair is the resource consumed to perform one quantum
  teleportation.  Simulations that include both the physical
  interactions and the extensive classical messaging confirm that
  Dijkstra's algorithm works well in a quantum context.  Simulating
  about three hundred heterogeneous paths, comparing our path cost and
  the total work along the path gives a coefficient of determination
  of 0.88 or better.
\end{abstract}





\section{Introduction}
\label{sec:intro}

A routing algorithm chooses a path on a graph, and consists of two
parts: a definition for the cost of a single link, and a function for
calculating the cost of a path based on those link costs, allowing us
to extend a single point-to-point channel to a richer network.
Dijkstra's Shortest Path First algorithm, for example, takes a simple
scalar cost for each link and treats the sum of link costs as a cost
for a candidate path~\cite{dijkstra1959ntp}.  The emerging field of
quantum communication has, to date, experimentally demonstrated the
basic principles of entangled quantum
networking~\cite{Chin-WenChou06012007,kimble08:_quant_internet,reichle2006ept,tashima:PhysRevLett.105.210503,zhao2003ere},
and laid the theoretical foundations of creating long-distance,
high-quality
entanglement~\cite{bennett:teleportation,briegel98:_quant_repeater,lloyd2004iqi},
but topologically has considered primarily channels and linear
networks, leaving us with an urgent need for a path selection
mechanism as quantum networks develop.

Quantum key distribution (QKD) is probably the most prominent use of
quantum communication, and commercial products are
available~\cite{bennett:bb84,dodson2009updatingQKD,lo:qkd-review}.  In
QKD, quantum effects (and a large dose of classical statistics) are
used to detect the presence of an eavesdropper on the channel.  QKD
generates streams of shared, secret, random bits, which can then be
used to key a cryptographic session, such as an IPsec
tunnel~\cite{elliott:qkd-net,mink09:_qkd_and_ipsec,nagayama09:_ike_for_ipsec_with_qkd,peev:secoqc}.
To extend QKD over distances exceeding a certain limit (a few hundred
km in telecom fiber today) or to create networks without direct links
between all pairs of nodes, we can use trusted relay nodes and direct
optical switching.  Such metropolitan-area QKD networks have already
been demonstrated in Boston, Vienna, Geneva, Tokyo, Hefei, and other
cities~\cite{dodson2009updatingQKD,elliott:qkd-net,peev:secoqc,chen2010metropolitan}.

Long-distance QKD without the trusted relay nodes, as well as most
other distributed quantum applications, can be achieved through the
use of {\em entanglement}~\cite{alleaume:njp-qkd,ekert1991qcb}
(entanglement and other italicized vocabulary will be explained in
more depth in Sec.~\ref{sec:q-c-diffs}).  Physical links can create
entanglement over short distances; in multi-hop networks, many
short-distance entangled states are used to forge the end-to-end
entangled states consumed by applications. Complex, irregular network
topologies with hops of varying length and quality, from quantum LANs
through MANs to WANs, will result as quantum systems grow.  The
devices that will bridge long distances and route connections through
networks are called \emph{quantum
  repeaters}~\cite{briegel98:_quant_repeater}.

Entangled states have uses beyond QKD.  The quantum communications
toolkit includes such building blocks as quantum Byzantine agreement,
distributed leader election, secure multi-party communication, and
distributed arithmetic for building large-scale quantum
applications~\cite{ben-or2005fast,buhrman03:_dist_qc,dhondt05:_dist-qc,tani05:_quant_leader_elect}.

\begin{figure}[h]
\begin{center}
\includegraphics[width=8cm]{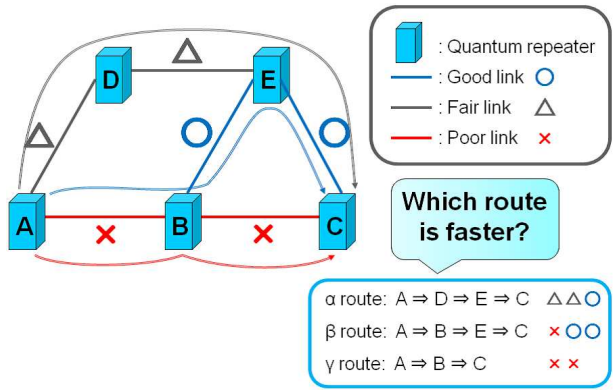}
\caption{Path selection is a critical problem in all networks; in
  quantum networks, due to the delicacy of quantum information, it may
  determine whether or not a connection can be made successfully.}
\label{fig:routing-example}
\end{center}
\end{figure}

Thus, heterogeneous quantum networks hover in our future, but
substantial theoretical and engineering hurdles remain to be cleared.
The first, theoretical studies on entangled quantum networks focused
primarily on an abstract model consisting of a linear chain of
repeaters, with a power of two number of hops of identical length and
quality~\cite{briegel98:_quant_repeater,dur2007epa}.  Recent
work~\cite{PhysRevLett.104.180503,PhysRevA.79.032325,munro2010quantum}
has targeted more realistic chains of repeaters, relaxing those
constraints.  Here, we analyze the behavior of more complex network
topologies, as in Fig.~\ref{fig:routing-example}. In a network of
heterogeneous links and irregular topology, path selection affects
both the performance of individual connections and global network
load.  {\em Purify-and-swap} quantum repeaters are a theoretically
well-developed approach, and are strong candidates for field
deployment, and we use them as our subject.

In this paper, we apply Dijkstra's algorithm for ranking candidate
paths to quantum repeater networks~\cite{dijkstra1959ntp}.  We
evaluate our algorithm for its ability to select a path that both
maximizes the throughput of end-to-end connections and minimizes
global work for this class of repeater. Detailed simulations of both
the physical interactions and the classical messaging confirm
reasonable agreement between the calculated path cost and the expected
throughput.

Our proposed link cost is the inverse of the throughput of the link,
measured in \emph{Bell pairs} per second of a particular {\em
  fidelity}.  Other candidates for link cost, including lower-level
metrics such as the number of laser pulses and quantum measurement
operations, are found to be useful for evaluating the total work
actually consumed on a path, but are poor metrics for prioritizing a
link because they reflect the physical link characteristics but not
the system factors that are equally important influences on end-to-end
performance.

We present the results of two sets of simulations of various paths
using four different qualities of links.  The first set of forty-six
paths vary in length from one to nine hops, while the second set
covers 256 link combinations in four-hop paths.  Across both data
sets, the coefficient of determination is 0.88 or better between the
path cost and the total work performed (counted as the number of
\emph{quantum measurements} performed along the whole path),
supporting our choice of link cost and the effectiveness of Dijkstra
for this type of quantum network.  Comparing the results of pairs of
simulations, the path with the lower cost also has higher throughput
in more than 80\% of all tested cases.  We demonstrate that, in direct
analogy to classical networks, the performance of a quantum path will
be limited by the throughput of the \emph{bottleneck link}, while
total work is a function of both the path length and the quality of
all the links.

To build a complete argument, we discuss the differences between
quantum and classical networks and the difficulties encountered
(Sec.~\ref{sec:q-c-diffs}).  After defining the problems in path
selection and proposing several solutions (Sec.~\ref{sec:problems}),
we evaluate those solutions via simulations that answer a series of
specific questions about the behavior of quantum repeater networks
(Sec.~\ref{sec:sims}).

\section{Difficulties: Differences Between Quantum and Classical
  Networks}
\label{sec:q-c-diffs}

So far in this paper, we have discussed quantum communication
technologies and outlined where we apply classical techniques to the
quantum problems.  However, we have not specifically articulated the
fundamental differences that make the merger of classical and quantum
networking concepts less than straightforward.  There are both
theoretical results and practical reasons for believing that the
answers to our questions require thought, rather than simply asserting
that classical and quantum networks can use the same solutions.

The differences stem from several sources: the engineering
difficulties of creating and protecting quantum states and the
real-time decay of quantum information; the impact of (probabilistic)
photon loss; and the fundamentally probabilistic nature of some
quantum operations.  These issues manifest themselves both locally and
globally, requiring in some cases additional classical messages to be
exchanged, further exacerbating the problems.

Perhaps most importantly, we cannot make independent copies of quantum
information, a fact known as the {\em no cloning
  theorem}~\cite{wootters:no-cloning}.  In classical systems, data is
copied as it is transmitted; if the data is lost, retransmission is
done with no harm beyond a performance penalty.  With only a single
copy of quantum data, we are forced to treat that data with extra
care.  Generally, quantum networks build generic states which are then
used to {\em teleport} precious data~\cite{bennett:teleportation}.  To
understand teleportation, we must first discuss entanglement; after
teleportation and its uses, we can complete the role of quantum
repeater by discussing imperfect quantum states.

\subsection{Entangled States}

Quantum communication involves the sending and receiving of quantum
information, encoded in either the state of an individual photon, or a
quantum state of a stronger optical or microwave pulse.  The state may
be sent through a waveguide such as an optical fiber, or through free
space, even via
satellite~\cite{tashima:PhysRevLett.105.210503,lo:qkd-review,peev:secoqc,fedrizzi2009high,scheidl2009feasibility,villoresi08:_space-quantum}.
A transmitted state may stand alone, or be part of a larger, {\em
  entangled} quantum state.  

{\em Entanglement}, which Einstein referred to as ``spooky action at a
distance'', is the condition of a quantum state in which the states of
subsystems are not independent: operations on one part of the system
can affect the state of other parts, regardless of physical distance.
This effect does not violate the principle of relativity, because it
cannot be used to transmit information instantaneously;
\emph{interpreting the meaning} of the distant parts requires the use
of classical information acquired during the local quantum operations.

That classical information about the quantum states is acquired via
{\em measurement} of the states.  Measuring a single quantum bit
(\emph{qubit}) produces one classical bit whose value depends on the
details of the quantum state, and destroys any entanglement of the
qubit with others.  The actual measurement mechanism is of course
technology-dependent, but often involves interferometric optical
setups with very sensitive photodetectors, allowing polarization or
other characteristics of an optical state to be determined, detecting
a single photon or measuring the direction of spin of a single
electron.  For the quantum states of interest here, the classical
information gained is guaranteed by quantum mechanics to be random,
and must be transmitted to the remote site, which of course can only
be done at the speed of light.

The particular type of entanglement we require is called a \emph{Bell
  pair}.  Although entanglement is location independent, when we refer
to a Bell pair in this paper, we mean a geographically distributed
Bell pair consisting of one qubit at one location entangled with a
qubit at another location.  Entanglement comes in many forms and can
involve more than two parties, but in this paper we focus on the
two-party Bell pairs both as intermediate resources and the end goal
of the communication session delivered to the application layer.

Quantum communication systems use entanglement in various ways, and
for different purposes.  Some uses require that the end nodes store
quantum data in memory for extended periods of time (many round-trip
times); others utilize quantum states of light, but measure the
state directly upon receipt without storing, converting the quantum
state to a classical value.  One of the greatest challenges in quantum
communication is dealing with memory lifetimes that are low compared
to round-trip latencies.

\subsection{Teleportation and Entanglement Swapping}

To teleport a qubit, we begin with a Bell pair with one end held at
our source and the other held at the destination.  Local entangling
operations are performed on the data qubit and the source-end Bell
pair qubit, then both are measured, giving two classical bits of
information.  Those two classical bits are transmitted to the
destination, which then performs local quantum operations on its
Bell-pair qubit, dependent on the values of those bits.  This
recreates the original data qubit at the destination.

\begin{figure}[h]
\begin{center}
\includegraphics[width=8cm]{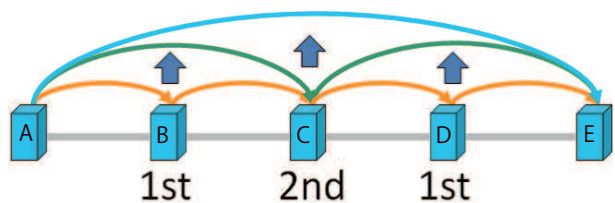}
\caption{Symmetric nested entanglement swapping, splicing
  shorter-distance entangled pairs to create longer-distance
  entanglement, is accomplished easily in $n$ steps for $2^n$ hops of
  the same quality.  The legend and arrows show the order in which
  entanglement swapping is performed.  The arcs represent entanglement
  over different distances, created via the swapping of shorter
  entangled pairs.}
\label{fig:sym-swap}
\end{center}
\end{figure}

Entanglement swapping (Fig.~\ref{fig:sym-swap}) splices two
short-distance Bell pairs into one longer-distance Bell pair: in
Fig.~\ref{fig:sym-swap} $(A\leftrightarrow B) + (B\leftrightarrow C)
\Rightarrow (A\leftrightarrow C)$ combines the two left gold arcs to
form the green one above.  Swapping is a form of teleportation; it can
be viewed as using the $(B\leftrightarrow C)$ Bell pair to teleport
the $B$ end of the $(A\leftrightarrow B)$ pair to $C$.  Classical
information must be sent to both ends of the new entangled connection.
Managed carefully, each swapping step can double the span of our
entangled Bell pair.

\subsection{Imperfect Quantum States}

In real systems, Bell pairs are imperfect, and have a characteristic
known as \emph{fidelity}, which represents the accuracy of our
knowledge about the quantum state, with $F = 1.0$ representing
perfection.  Mathematically, the fidelity is the amount of overlap
between the state we intended to create and an average over an
ensemble of states we actually created, either experimentally or in
simulation, and is defined as $F = \langle\Psi^+|\rho|\Psi^+\rangle$,
where $\rho$ is the {\em density matrix} representation of the state
we have, and $|\Psi^+\rangle$ is the Bell state we are trying to
create.  In experiments, fidelity is determined statistically using
large numbers of repetitions, creating a quantum state and measuring
it to see if the state was properly made and stored.  In simulations,
various imperfections are numerically introduced on the state
creation, processing, or storage to mimic the experimental procedure.
By simulating smaller or larger imperfections, we can guide the
experimental focus toward the most critical problems, and establish
fidelity targets which experimentalists must meet for systems to be
operationally viable.  In network operations, the results of a set of
experiments run during link initialization will be used to track
expected behavior of the system.

Purification takes two imperfect ($F < 1.0$) Bell pairs that span one
or more hops and attempts to create one higher-fidelity Bell pair with
fidelity $F'$, ($F < F' < 1.0$).  The biggest disadvantage of
purification is that it is probabilistic, requiring two-way classical
communication, tying up buffer memory resources and reducing
throughput.

\subsection{Quantum Repeaters}

The devices that create Bell pairs over a distance are called
\emph{quantum repeaters}, building on the concept of teleportation.
Because teleportation is imperfect, some form of error control is
necessary.  In this paper, we focus on the form of repeater we will
call the \emph{purify-and-swap} approach.  Purify-and-swap repeaters,
as the name suggests, perform two main functions: \emph{entanglement
  swapping} to lengthen connections and \emph{purification} to manage
errors~\cite{briegel98:_quant_repeater,dur2007epa}.  Bell pairs are
consumed during the course of teleportation, purification and
entanglement swapping; thus, the primary job of a repeater is to
continually produce new ones.  From Bell pairs spanning shorter
distances, we can create end-to-end Bell pairs, which are then given
to applications.  ``Throughput'' can be defined as Bell pairs per
second of a given fidelity, for a given path.

We have defined a protocol stack for purify-and-swap
repeaters~\cite{van-meter07:banded-repeater-ton}, consisting of three
layers: Entanglement Control (EC) at the link level, and Purification
Control (PC) and Entanglement Swapping Control (ESC).  The latter two
operate at arbitrary distance.  They can be composed
recursively~\cite{touch2006recursive} to build high-fidelity Bell
pairs in a nested fashion across any nodes in a network, but doing so
requests resources which depend critically on the route chosen in the
network.  The operations of these protocol stack elements are
therefore the ones we vary in the present work.

As we shall see in Sec.~\ref{sec:1hop}, the low probability of
entanglement success requires that link-layer operations be
acknowledged.  Coupled with very limited buffering resources, this
results in behavior that differs from most classical networks.
Quantum memory is very expensive, simply because of the (usually
cryogenic) hardware needed to maintain quantum states for timescales
comparable to photon time-of-flight, and to connect that hardware to
photons.  The fidelity of data in quantum memory in most technologies
decays very quickly, adding a hard real-time component to
communication, or a demand for strong error correction.  This decay
may make congestion control a different, even more severe, problem
than it is in classical networks.

For a classical path consisting of a series of identical links,
throughput (at least in the ideal, sustained case) can match the
throughput of a link, independent of the number of hops.  However, the
maximum throughput of a symmetric chain of purify-and-swap repeaters
declines polynomially with length, due to the need for additional
purification~\cite{dur2007epa}.  Thus, it
\emph{may} be desirable to assign a path cost that grows more than
additively.

When operated to deliver Bell pairs above a specified fidelity
threshold (typically related to the fidelity threshold for executing
quantum error correction), simulations show a stair-step decline in
throughput versus hop count~\cite{van-meter07:banded-repeater-ton}.
The total work performed grows linearly as the throughput holds
steady, then increases suddenly when a ``stair'' is crossed (see
Figs.~\ref{fig:1hopdata} and ~\ref{fig:t-p-m} in Sec.~\ref{sec:sims}).
Likewise, dealing with non-power-of-two numbers of hops will affect
throughput and total work performed in hard-to-predict fashion, due to
changes in the swapping and purification patterns.

These principles have been applied within designs for system-area
networks, as well as wide-area networks.  Interconnects are critical
components of quantum computing
systems~\cite{copsey:q-com-cost,isailovic06:_interconnect,1330522,oskin:quantum-wires};
even some laboratory-sized experiments, much like classical
multicomputers, will have multiple levels of interconnect, with
different qualities and speeds of
connection~\cite{van-meter10:dist_arch_ijqi}.  These and other
large-scale systems will use purified Bell states for data movement.
Purification forms a large chunk of the work in these systems as well
as in larger networks, and its demands are a commonly-used metric for
evaluating systems.

Recent results suggest that using quantum error
correction instead of purification opens up new operational approaches
more akin to the hop-by-hop behavior of packet-switched networks, and
may prove to be more efficient in many
cases~\cite{PhysRevLett.104.180503,PhysRevA.79.032325,munro2010quantum}.
These new approaches reduce the demands on memory lifetime and limit
the number of return messages required.  However, the memory resources
required are substantial, far exceeding what will be practical in the
near future.  The earliest deployments of repeater networks, and
perhaps the first commercial products, will likely use the
purify-and-swap approach.  Thus, we reserve analysis of our algorithm
for other types of networks (and heterogeneous combinations of network
types) for future work.

Although some authors have begun speculating about a ``quantum
Internet'', the issues of complex, entangled networks, and
heterogeneity at any level above the physical exchange of qubits from
one medium to another, are largely open
problems~\cite{kimble08:_quant_internet,lloyd2004iqi}.



\section{Problems and Solutions}
\label{sec:problems}

\comment{This needs sharpening.}  Our goal for this paper is to
develop and analyze a routing algorithm for heterogeneous quantum
repeater networks, as shown in Fig.~\ref{fig:routing-example}.  Our
metric for success, therefore, is agreement between the prospective
functioning of a network (represented by detailed
simulations) and an easy-to-calculate algorithmic cost: does our
algorithm allow us to make effective choices?


As noted above, to develop a routing algorithm, we need a definition
for the cost to use a link, a function to calculate a path cost based
on a set of link costs, and a goal for the algorithm (e.g., a metric
for deciding if the algorithm meets our needs).  More precisely, we
set as our problems:

\begin{description}
\item[{\bf PS.1}] Choose a goal for the routing algorithm;
\item[{\bf PS.2}] for a quantum link, identify the characteristics of
  interest for routing, and reduce them to one number or a small set
  of numbers that represent the link cost; and
\item[{\bf PS.3}] for a path (an ordered set of links, with associated
  costs), define a function that gives a path cost.
\end{description}

To solve these problems, we evaluate the following potential
solutions:

{\bf PS.1: Goal} As the goal for the routing algorithm itself, we
choose \emph{minimizing work along the path}, with attention to the
secondary goal of selecting the highest-throughput single path between
the defined communication endpoints, measured in Bell pairs per second
of the target fidelity.  We propose a goal for the \emph{system} of
delivering Bell pairs useful for teleportation, with a target fidelity
of $F = 0.98$.  The phenomena presented here are independent of the
exact value chosen, but this value will be adequate for various uses,
and allows fairly direct comparison to our prior
work~\cite{van-meter07:banded-repeater-ton}.

Assessing the work for a particular path is not a simple problem.
Intuitively, we want our measure of work to reflect use of some scarce
resource.  We evaluate two candidates discussed as key functions in
repeaters above: {\em total measurement operations} and {\em total
  pulse count}.

Without some reasonable idea of
global traffic (which we don't yet have)~\footnote{However, for
  heterogeneous traffic matrices, a gravity model might be a
  reasonable first guess~\cite{medina2002traffic}, as the human
  patterns driving quantum communication are expected to be similar to
  those for classical communication.}, evaluation of global success is
difficult, so as a first step we are evaluating the correlation
between total work on the path, achieved connection throughput and our
definition of path cost.

{\bf PS.2: Link Cost} By analogy with the ``transceiver time''
definition often used with OSPF, we are exploring several link cost
definitions.  Simply assessing the clock speed at which laser pulses
can be emitted is clearly an inadequate link cost metric in a system
where the fidelity of the output state is important and the
probability of success depends on the characteristics of the link.
Prior to execution of our experiments, differing intuitions led one
author to suggest Bell pair generation time and another to suggest the
number of measurements; eventually we had a list of five candidates:

\begin{description}
\item[{\bf Loss}] the loss in the channel, in decibels;
\item[{\bf InvTrans}] the inverse of the transmittance of the channel
  ($1/T$, where the transmittance $T$ is the percentage of photons
  received through the path);
\item[{\bf Pulse}] the number of laser pulses used to
  create an entangled Bell pair of a high fidelity over a single hop,
  corresponding to the number of uses of the transmitter (each
  fixed-time);
\item[{\bf Meas}] the number of measurement operations used to create an
  entangled Bell pair of a high fidelity over a single hop (this
  differs from the above because (a) some pulses are discarded rather
  than accepted when receiving qubit resources are busy, and (b)
  measurements are used in entanglement swapping and purification);
  and
\item[{\bf BellGenT}] the inverse of the throughput of the single link, when
  run as a single-hop system, measured in seconds per Bell pair.
\end{description}

This list can be divided into two groups: the first two candidates are
simply physical characteristics of the link that can be measured
easily, while the latter three require simulation or monitoring of the
link to determine.  The first two differ by a logarithmic factor;
InvTrans corresponds to addition while Loss in dB corresponds to
multiplication of cost when placed in a Dijkstra context.  While at
first glance it may appear desirable to have such an easily-determined
link cost, by tying the cost so directly to the physical mechanism,
the definition may not transfer well when links of heterogeneous
physical technologies are involved.

Pulse and Meas seem to correspond most closely to the ``transceiver
time'' definition, but BellGenT incorporates system factors and may
give more accurate estimates for little additional complexity.
(Although the technical details are very different, these could be
considered roughly analogous to the raw transceiver rate, the
throughput of a flow-controlled link, and the throughput of a reliable
link-layer protocol, in terms of the increasing functionality
present.)

{\bf PS.3: Path Cost Function}
The principle hypothesis of our paper is that Dijkstra's algorithm can
be used as-is, with an appropriate choice of link cost.  More
formally,
\begin{equation}
C_{path} = \sum_i{c_i}, i \in \{ P \},
\end{equation}
where $\{ P \}$ is the set of links in a path and $c_i$ is the cost
for link $i$.  When it is necessary to distinguish among the link cost
candidates, we will refer to them as Dijkstra/BellGenT and similarly.


\section{Simulation and Results}
\label{sec:sims}

Our goal is to determine the range of conditions under which Dijkstra
correctly selects the lowest-work path and highest-throughput path,
and when it selects some reasonable approximation.  We also wish to
articulate the conditions under which lowest work and highest
throughput are not the same path and when the algorithm does not pick
the lowest-work path.  We wish to examine whether we consider those to
be acceptable cases, or if the algorithm is ``failing''.  This can be
achieved by creating a set of candidate paths and comparing the
ordering established by Dijkstra/BellGenT and the ordering according
to simulation of the whole path.

In this section, we begin by asking a set of questions about the
behavior of heterogeneous paths (\ref{sec:questions}).  Next, we
describe our simulator, proposed hardware configuration and single-hop
simulation results, which both set the Dijkstra parameters and allow
us to evaluate our link cost candidates (\ref{sec:1hop}).  After
enumerating a set of interesting path candidates (\ref{sec:path-can}),
we can answer our questions (\ref{sec:answering}), and use those
answers to solve the research problems posed in
Sec.~\ref{sec:problems} (\ref{sec:solving}).

\subsection{The Behavior Questions}
\label{sec:questions}

We use the simulator to assess the behavior of systems that are too
complex to solve analytically and cannot yet be built and measured
directly.  Such simulation results will help to guide the development
of actual hardware.  We can pose a series of specific questions that
will help us understand how heterogeneous paths will behave:

\begin{enumerate}
\item\label{q:hopno} How does the \emph{number} of hops in a path
  affect throughput and total work?
  As a specific case, do the throughputs of $2^n-1$, $2^n$, and
  $2^n+1$-hop paths vary?
\item\label{q:bottleneck} How does the number of \emph{weak} links
  matter?  Does the introduction of a single weak link become a
  \emph{bottleneck}, as in classical networks?  Does adding a second
  or third weak link further reduce throughput?
\item\label{q:bpos} How does the \emph{position} of weak links in the
  chain affect throughput?  Does a weak link at the beginning, in the
  middle, or at the end differ?
\item\label{q:mis-order} Under what circumstances will the path cost
  mis-order candidate paths with respect to throughput or total work?
\end{enumerate}

For classical systems, we know the answers: although cost increases,
(1) the length of the path (in hops) has no effect on the
(theoretical) throughput, and (2) the (theoretical) throughput of a
path is capped by the throughput of the bottleneck link(s).  Question
(3) is answered in the negative: ordering does not (or should not)
matter, and ordering of path segments should always agree with
ordering of the full paths.  Question (4) is open-ended, but there are
known cases where a lower-performance path can be selected in the
service of a larger global goal.  The answers to these questions are
presented in Sec.~\ref{sec:solving}.

\subsection{Simulated Hardware and Link Costs}
\label{sec:1hop}

All of the simulations presented in this paper model the \emph{qubus}
physical entanglement
mechanism~\cite{ladd06:_hybrid_cqed,van-loock06:_hybrid_quant_repeater}.
Qubus uses a strong light pulse to entangle two distant, static qubit
memories through a fiber or other waveguide.  Alternative approaches
use single photons~\cite{childress2006ftq} or a small number of
photons~\cite{munro:PhysRevLett.101.040502}.  In the qubus system, a
strong laser pulse first interacts with a physical qubit at the
transmitter, such as an atom or a quantum dot held in a cavity,
resulting in a small, nonlinear shift in the state of the laser pulse,
depending on the state of the qubit.  The pulse is sent through a
waveguide to the receiver, where it undergoes a similar interaction
with another qubit.  The pulse is then \emph{measured} using a
technique known as homodyne measurement, giving a classical result.
About 40\% of the time, we get a result that tells us only the parity
of the two qubits and leaves them entangled.

We use the same simulator as in our prior work, with extensions to
support the heterogeneous
paths~\cite{van-meter07:banded-repeater-ton,ladd06:_hybrid_cqed,munro:PhysRevLett.101.040502}.
The simulator was developed for modeling the quantum-level behavior of
a cavity QED system for qubus and two other physical layer candidates,
and uses the purification mechanism of Dehaene {\em et
  al.}~\cite{dehaene2003lpp}.  The necessary classical messaging is
carefully modeled.  The simulator consists of about 11,000 commented
lines of C++, and the production runs of the simulations presented
here consumed about one hundred hours of CPU time on 2.2GHz AMD
Opteron CPUs.  The simulator is based on well-understood physical
equations and is experimentally validated at the lowest levels; as
experience with larger-scale quantum networks develops, we plan to
continue tracking the agreement of theory, simulation, and experiment.
We believe the prospective agreement between a real-world network and
our simulations will be more than adequate for the purposes of this
paper, but the detailed arguments are largely about physics, and are
beyond the scope of this paper.

\begin{figure}[h]
\begin{center}
\includegraphics[width=8cm]{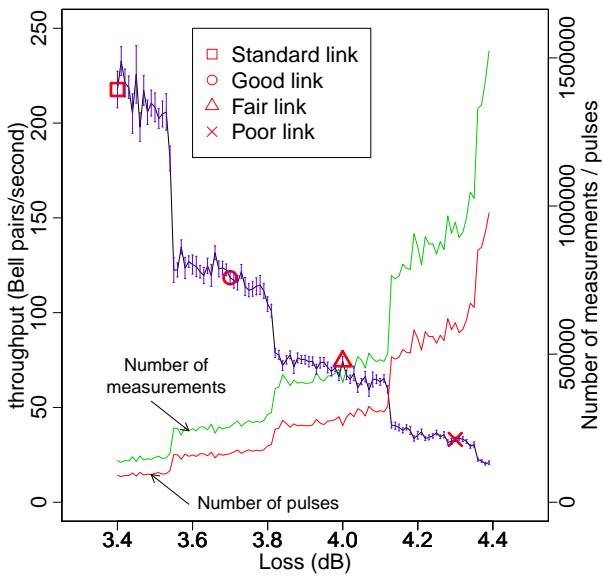}
\caption{Single-hop simulated throughput versus loss using the
  parameters in Table~\protect\ref{tab:1hop_environment}, used to
  define linkcost.  The four link types chosen for simulation of more
  complex paths are marked on the black/blue curve with the
  corresponding symbols.  The stair-step behavior is due to the
  increasing number of purification rounds as the initial fidelity
  declines due to increasing loss.  The details of this curve are very
  specific to the qubus technology we simulate, but the principles are
  general.}
\label{fig:1hopdata}
\end{center}
\end{figure}

Figure~\ref{fig:1hopdata} shows the results of simulating a single
qubus hop, with parameters as in Table~\ref{tab:1hop_environment};
additional details of the hardware configuration are the same as in
Refs.~\cite{van-meter07:banded-repeater-ton,ladd06:_hybrid_cqed,munro:PhysRevLett.101.040502}.
The original qubus mechanism is very sensitive to loss, but works well
in low-loss situations.  This form of qubus repeater fails to work for
losses greater than about 5.5dB from transmitter qubit to receiver
qubit, limiting hop length to about 30km over high-quality optical
fiber at telecom wavelengths.  Other types of physical link, including
variants of the qubus mechanism~\cite{munro:PhysRevLett.101.040502},
will work over longer distances and with higher fidelity.  By choosing
to simulate basic qubus links, we can see very clearly the impact of
purification and low-fidelity entanglement.  As future work, we plan
to confirm the behavior of Dijkstra with other physical link types.

For the multi-hop simulations, we choose four specific points as
example links, as shown in Table~\ref{tab:links}.  The four link types
chosen are marked in the figure with the corresponding symbols.  The
terms ``standard'', ``good'', ``fair'', and ``poor'' are relative to
this simulation only, not indicative of all possible physical quantum
link types.  As can be seen from this table and
Figure~\ref{fig:1hopdata}, even small differences in loss have a large
and uneven impact on throughput, suggesting that not only is the
utility of the proposed link costs Loss and InvTrans rather
technology-specific, but even in the isolated case of qubus they may
be poor metrics.

\begin{table}[h]
\begin{center}
\caption{Link cost simulation parameters}
\label{tab:1hop_environment}
\begin{tabular}{|c|c|}
\hline
\multicolumn{2}{|c|}{Qubus Quantum Repeater} \\
\hline \hline
Number of qubits & 25 transmitter, \\
per repeater link connection & 25 receiver \\
\hline
Number of qubits & \multirow{2}{*}{200 qubits}\\
teleported (length of simulation)& \\
Final target fidelity & $F \ge 0.98$ \\
\hline
\hline
\multicolumn{2}{|c|}{Optical Fiber} \\
\hline
\hline
Length & 20 km\\
\hline
Signal loss & 3.4$\sim$4.4 dB/20 km\\
\hline
\end{tabular}
\end{center}
\end{table}

\begin{table*}
\begin{center}
\caption{The link configurations and five candidates for link
  cost.  The cost candidates to the left of the first
  double line are physical characteristics of the channel, while the
  others are the results of single-hop simulations.  Simulation
  parameters are as in Table~\protect\ref{tab:1hop_environment}.}
\label{tab:links}
\begin{tabular}{|l|r|r||r|r|r||r|r|r||r|r|}\hline
Link & Loss & InvTrans & Pulse & per-t'port & (norm) & Meas &
per-t'port & (norm) & T'put & BellGenT\\\hline
Standard ($\Box$) & 3.4dB & 2.19 & 90441 & 452 & 1 & 140519 & 702 & 1 & 217.7 & 1 \\
Good ($\bigcirc$) & 3.7dB & 2.34 & 163628 & 818 & 1.80 & 254691 & 1237 & 1.76 & 118.4 & 1.83  \\
Fair ($\triangle$) & 4.0dB & 2.51 & 258852 & 1294 & 2.86 & 404117 &
2020 & 2.87 & 74.3 & 2.93  \\
Poor ($\times$) & 4.3dB & 2.69 & 606278 & 3031 & 6.70 & 945247 & 4276 & 6.72 & 33.1 & 6.57 \\
\hline
\end{tabular}
\end{center}
\end{table*}

Although Bell pairs are symmetric and can be used to teleport data in
either direction, at the physical level, links are naturally dependent
on a transmitter and a receiver.  Receivers dynamically select a qubit
to attempt to entangle, based on availability at the time of pulse
reception, much as any network interface chooses a buffer in which to
place an incoming packet.  The inherent round-trip latency in
receiving acknowledgement of entanglement success or failure (the
Entanglement Control (EC) protocol layer) means that the qubits at the
transmitter must spend long periods of time in which they are
nominally marked as busy, but no useful forward progress can be be
made.  Thus, links with the same physical loss but different numbers
of qubits at the transmitter and receiver will behave differently, as
shown in Table~\ref{tab:link-qubits}.

Although the throughput varies by a factor of two for different
configurations, the number of pulses and number of measurements to
teleport 200 qubits does not vary significantly except for the clearly
misconfigured 100/25 case.  For the conditions simulated, about 450
pulses and 700 measurements (including basic entanglement, swapping
and purification) are required to teleport one qubit.  Although these
two measures are arguably more direct representations of cost, this
important throughput difference affects our notion of a preferred
path.

Reasoning about the behavior of only a single hop and confirming the
results via simulation allow us easily to eliminate two of our
prospective link cost candidates (Loss and InvTrans).  Two others
(Pulse and Meas) remain good measures of total work, but the results
shown in Tab.~\ref{tab:link-qubits} cast doubt on their viability as
candidates for link cost when the goal is to achieve high throughput.
Thus, \emph{we settle on BellGenT as our link cost metric}.

\begin{table}
\begin{center}
  \caption{Throughput depends on the number of available qubits at
    both transmitter and receiver.  Due to the round-trip latency in
    the Entanglement Control (EC) protocol for acknowledging
    successful entanglement, having more qubits at the transmitter
    boosts throughput.  Simulations are for a single hop with
    parameters as in Table~\protect\ref{tab:1hop_environment}.
    Throughput and confidence interval (std. dev.) are determined by a
    linear fit to teleportation completion times.}
\label{tab:link-qubits}
\begin{tabular}{|c|c|c|c|c|}\hline
Xmtr & Rcvr & Throughput & Pulses & Meas \\\hline
25 & 25 & $237\pm 8$ & 80587 & 125905 \\
25 & 50 & $213\pm 8$ & 92071 & 143582 \\
50 & 25 & $436\pm 14$ & 99708 & 142723 \\
50 & 50 & $456\pm 14$ & 89533 & 139506 \\
100 & 25 & $462\pm 22$ & 199240 & 138543 \\
100 & 50 & $956\pm 31$ & 92221 & 137828 \\
100 & 100 & $984\pm 32$ & 88694 & 138145 \\
\hline
\end{tabular}
\end{center}
\end{table}

\subsection{Simulated Path Candidates}
\label{sec:path-can}

The candidate paths we choose to examine are designed to answer the
questions in Sec.~\ref{sec:questions}.  We simulated forty-six paths
of one to nine hops in various patterns, as well as all $4^4 = 256$
four-hop combinations for the four chosen link types.  The forty-six
paths are enumerated along the bottom of Fig.~\ref{fig:t-p-m}.  We
have simulated many more paths with a variety of link conditions over
the course of this experiment, notably very long, homogeneous paths
(up to 2,048 hops).  The findings from other simulations do not
contradict the results presented in this paper, which were chosen to
clearly show the effects of interest.

\subsection{Answering Our Behavior Questions}
\label{sec:answering}

\begin{figure}[h]
\begin{center}
\includegraphics[width=8.5cm]{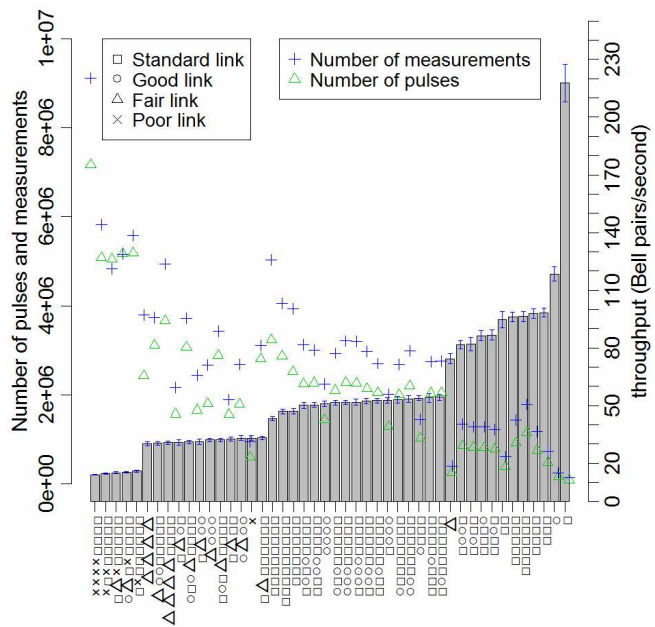}
\caption{Total number of pulses ($\bigtriangleup$) and measurements
  ($+$) for forty-six of our candidate paths (right scale).  The paths
  vary in length from one to nine hops.  They are ordered left to
  right according to ascending throughput, plotted using bars (left
  scale).  The legend below the graph shows the individual path
  configurations; $\Box$, $\bigcirc$, $\bigtriangleup$ and $\times$
  represent our standard, good, fair, and poor links, respectively.
  The stair-step behavior reflects increasing numbers of rounds of
  purification.}
\label{fig:t-p-m}
\end{center}
\end{figure}

\begin{figure}[h]
\begin{center}
\includegraphics[width=8cm]{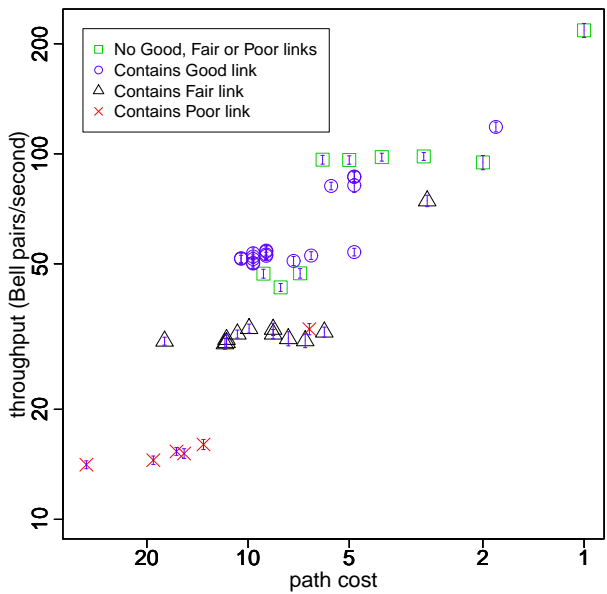}
\caption{Throughput versus BellGenT path cost for forty-six of our candidate
  paths. Each path is represented by the symbol for the weakest type
  of link in the path.  The clustering of each type of data point shows
  clearly that throughput is limited by the bottleneck link.  The
  length of the vertical bar (mostly contained within each symbol)
  shows the std. dev. of the throughput.}
\label{fig:first-results}
\end{center}
\end{figure}

\begin{figure}[h]
\begin{center}
\includegraphics[width=8cm]{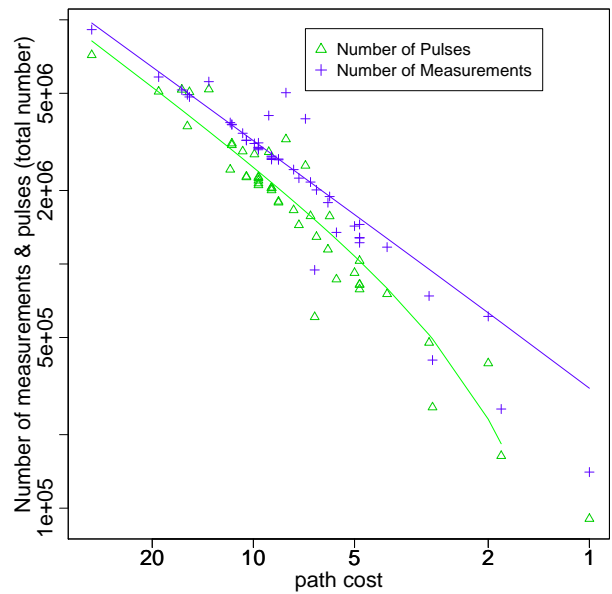}
  \caption{Total work, measured in Pulses ($\bigtriangleup$) and
    Measurements ($+$), versus BellGenT path cost for forty-six of our
    candidate paths.  The coefficient of determination of each linear
    fit is 0.88, showing that our path cost is a strong predictor of
    total work.}
\label{fig:total-v-path}
\end{center}
\end{figure}

\begin{figure}[h]
\begin{center}
\includegraphics[width=8cm]{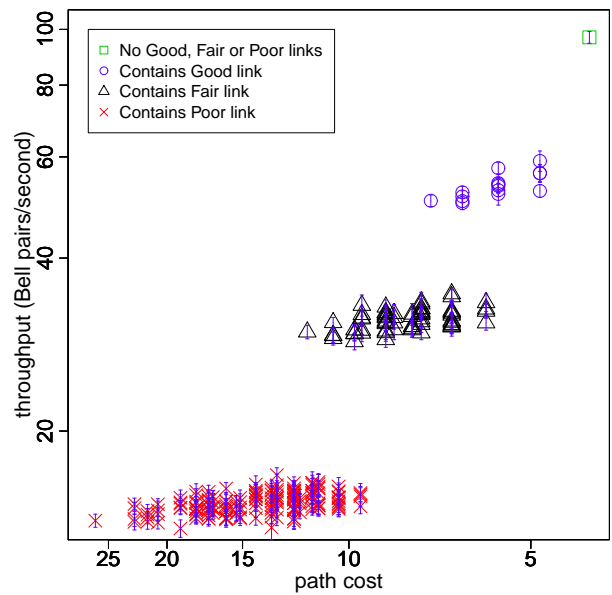}
\caption{Throughput versus BellGenT path cost for all 256 four-hop candidate
  paths. Each path is represented by the symbol for the weakest type
  of link in the path.  The clustering of each type of data point shows
  clearly that throughput is limited by the bottleneck link.}
\label{fig:thru-v-cost-4hop}
\end{center}
\end{figure}

\begin{figure}[h]
\begin{center}
\includegraphics[width=8cm]{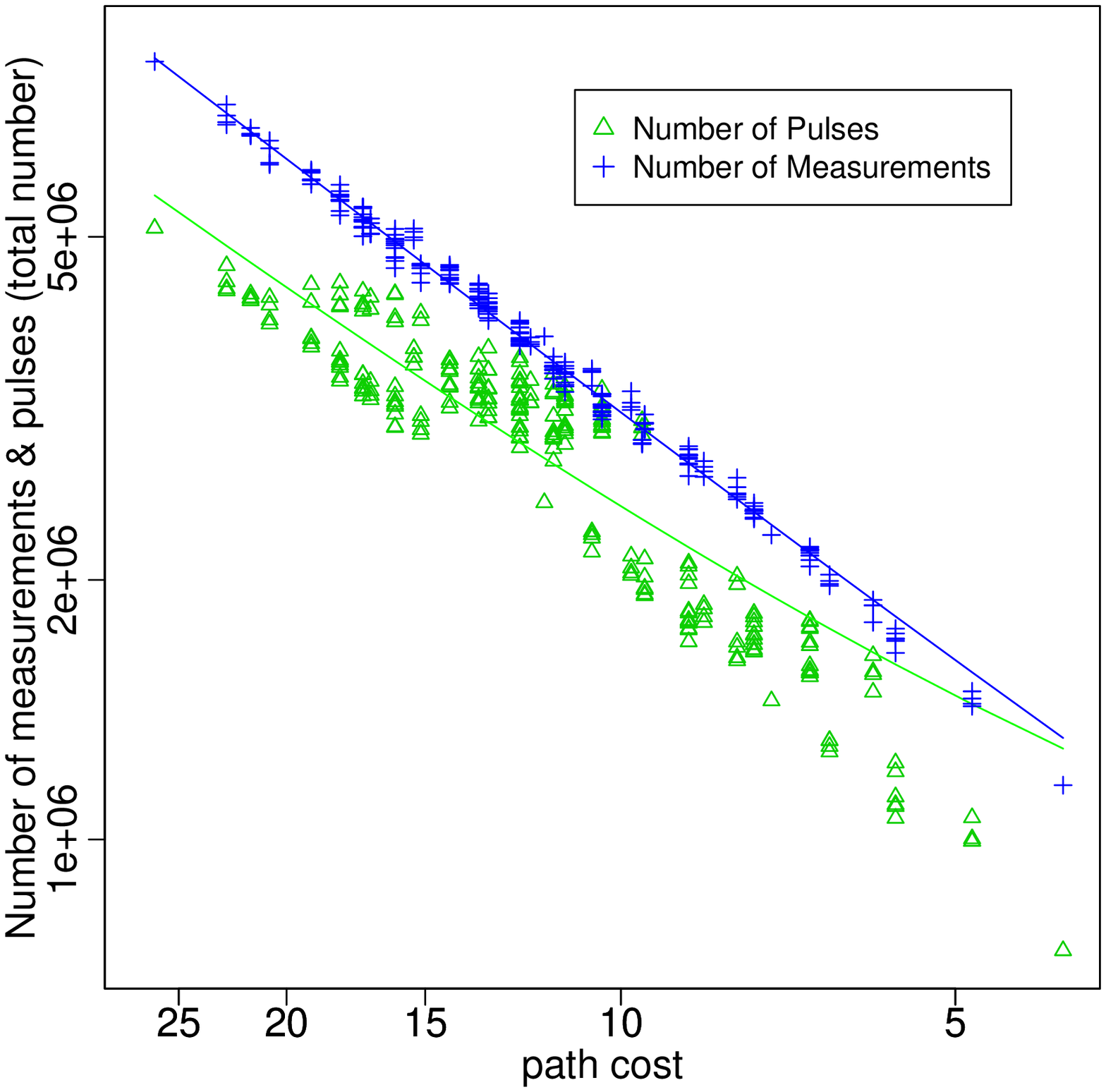}
\caption{Total work, measured in pulses ($\bigtriangleup$) and
  measurements ($+$), versus BellGenT path cost for all 256 four-hop
  candidate paths, with linear fits.  The coefficient of determination
  for the number of pulses is 0.81, and for the number of measurements
  is 0.99.}
\label{fig:total-v-path-4hop}
\end{center}
\end{figure}

Figures~\ref{fig:t-p-m}, \ref{fig:first-results} and
\ref{fig:total-v-path} show our simulation results for the paths.  In
Fig.~\ref{fig:t-p-m}, the throughput for each specific path can be
seen, as well as the two measures of total work, Pulse and Meas.  In
Fig.~\ref{fig:first-results}, throughput is plotted against the
Dijkstra-calculated path cost using BellGenT as our cost metric, and
in Fig.~\ref{fig:total-v-path} the total work measures are plotted
against calculated path cost.  Figures~\ref{fig:thru-v-cost-4hop} and
\ref{fig:total-v-path-4hop} plot the results for all 256 four-hop
paths we simulated.



Perhaps the most important factor to note about the behavior we see is
the discrete nature of changes to throughput, due to the discrete
nature of purification and entanglement swapping and our choice to
establish a particular threshold for final acceptance of an end-to-end
Bell pair.  In many circumstances, minor changes force an additional
round of purification, generally causing a 50\% reduction in
throughput.

As the \emph{number of hops} ({\bf Q.~\ref{q:hopno}}) increases, our
results show a stair-step phenomenon in throughput, but not
necessarily at powers of two: the decline of fidelity resulting in
additional rounds of purification does not move in concert with
entanglement swapping.  For Standard $\Box$ paths, we see steps at the
2nd and 7th hops.


Examining the symbols in Fig.~\ref{fig:first-results} shows clearly
the existence of a \emph{bottleneck link} phenomenon ({\bf
  Q.~\ref{q:bottleneck}}): with one low-quality link in the path, the
quality of the other links is almost irrelevant, as can be seen by the
clustering of each type of data point (e.g., $\bigtriangleup$ at about
30 Bell pairs/second in Fig.~\ref{fig:first-results} and the
concentration of $\times$ links at the left edge of
Fig.~\ref{fig:t-p-m}).  The various paths can largely, though not
entirely, be grouped according to the throughput of the slowest link
in the path.  The most eye-catching anomalies in
Fig.~\ref{fig:first-results} are the single $\bigcirc$,
$\bigtriangleup$ and $\times$ marks above and to the right of their
respective clusters.  These are the single-hop paths of the
corresponding link type, indicating that we do not have a pure
bottleneck phenomenon.  The bottleneck plus the polynomial decline in
performance as the number of hops grows work in combination determine
to the path performance.  In most cases for the longer paths, adding a
second link of the same quality as the bottleneck link does not result
in a statistically significant reduction in throughput, but
optimization of the path becomes more difficult.  In a few cases, our
optimization fails to find an acceptable pattern, and an additional
round of purification becomes necessary.  In particular, four-hop
paths with $\bigcirc$ as the bottleneck link(s) are split into two
plateaus at around 80 Bell pairs/second and 50 Bell pairs/second.
However, the total work shows strong correlation even for these cases.

Despite this general bottleneck behavior, the \emph{weak link
  position} ({\bf Q.~\ref{q:bpos}}) is a more subtle one.  Among our
simulations, we found a single case where the bottleneck position
produced different results.  With one Good link and three Standard
ones, having the Good link at the left end of the path resulted in
$53$ Bell pairs/second, whereas the other three paths produced $82$ to
$87$ Bell pairs/second.  The low-throughput case required an extra
round of purification before one entanglement swapping operation.  As
the path hovers near a threshold demanding an extra round of
purification, optimization of the path usage becomes more difficult;
three paths successfully did so, while the fourth didn't.  Thus, we
must answer that the weak link position \emph{may} have an impact on
throughput, depending on our ability to effectively use the path.

Dijkstra/BellGenT occasionally \emph{mis-orders} ({\bf
  Q.~\ref{q:mis-order}}) pairs of path candidates with respect to
throughput.  Comparing the 256 four-hop paths we simulated, there are
32,640 possible pairs, of which 1,230 had the same path cost.  Of the
pairs with different costs, $82.6\%$ of the time Dijkstra/BellGenT
chooses the higher-throughput path of the pair (the ``correct''
choice), and $17.4\%$ of the time it chooses the lower-throughput path
(the ``incorrect'' choice).  In only $5\%$ of those mis-ordering cases
was the difference in throughput more than $10\%$.  For the 46
variable-length paths, the rate of ``correct'' choices was similar,
$81.6\%$, but the impact of those choices is higher, with a throughput
penalty of $25\%$ or more in almost half of the ``incorrect'' cases.

The most remarkable case of incorrect ordering we found is the pair of
paths $\Box\times\Box\Box\Box\Box\Box\Box$ (throughput: $16.0\pm 0.51$
cost: 13.57) and
$\triangle\triangle\triangle\triangle\Box\Box\Box\Box$ (throughput:
$30.7\pm 0.77$ cost: 15.72).  The higher-cost path has twice the
throughput of the lower-cost one.  Examining Fig.~\ref{fig:t-p-m} shows
that the total work (in pulses or measurements) is quite similar.
This isolated case would suggest that the bottleneck Poor link might
warrant even a higher link cost than 6.57, however, in all of the
other cases we examined, it has worked well.

The performance and work are both dominated by purification costs.
The use of a given path must be optimized as a whole; our current
solution is brute force, trying a large number of possibilities.  In a
large majority of cases, optimization of that process for a path is
straightforward and robust.  However, a noticeable minority of cases
require delicate adjustments to the entanglement swapping settings.
Further automated optimization of this process may result in both
better performance and agreement between path cost and work.

For total work, we find a coefficient of determination of 0.88 for
both pulses and measurements for the forty-six variable-length paths.
For the 256 four-hop paths, we find a coefficient of 0.81 for pulses
and 0.99 for measurements.

We can summarize the behavior as follows: Performance is determined by
the number of rounds of purification used anywhere on the path; the
number of rounds is dominated by the bottleneck link.  Work, however,
is spread across the entire path in rough proportion to relative link
quality.

\subsection{Solving Our Problems}
\label{sec:solving}


%
With the simulator results and answers to our questions in hand, we
are now prepared to assess the solutions we proposed in
Sec.~\ref{sec:problems}, and determine whether Dijkstra/BellGenT
meets our goal of developing an acceptable routing algorithm for
quantum repeater networks.

We have seen that the highest-throughput path is also strongly
correlated to measures of work, including total pulse count and number
of measurements.  In some pathological cases, number of measurements
is a better metric, and across a broad range of cases it matches well
with our chosen link cost below.  Our solution to {\bf Path Selection
  Problem PS.1} is therefore: \emph{to minimize total work along the
  path, using number of measurements as the metric}.

We saw in Sec.~\ref{sec:1hop} that defining BellGenT as our link cost
metric is likely to better suit our purposes than Pulse, Meas, Loss,
or InvTrans, allowing us to propose a solution to {\bf Problem PS.2}:
\emph{seconds per Bell pair}.

Our simulation results presented in the previous subsections indicate
clearly that we can usually, if not always, correctly predict the
highest-throughput, lowest-work path from among a set of candidates.
Dijkstra/BellGenT solves {\bf Problem PS.3} well: \emph{path cost is
  the scalar sum of link costs}.

Intuitively, we can see why this works well, as confirmed by the
simulations: BellGenT is directly related to the number of
purification rounds needed, and hence the amount of work on a link.
Despite the concerns we expressed earlier about the nonlinear amount
of work as the number of hops grows, the predictive ability for
ordering paths remains strong.  Thus, we can assert that even as the
use of complex repeater networks evolves, and various traffic patterns
arise, Dijkstra/BellGenT likely will remain an effective choice.

\section{Conclusion}
\label{sec:conclusions}

Real-world deployment of networks of quantum repeaters will inevitably
be physically heterogeneous, with complex topologies and high- and
low-quality links and many possible paths through the networks, rather
than an idealized, homogeneous power-of-two number-of-hops.

In this paper, we have investigated critical problems in the use of
purify-and-swap repeater networks.  Our focus has been on path
selection, and the need for a routing algorithm.  An acceptable
routing algorithm must be an easy-to-calculate metric that reliably
chooses a reasonable, if not optimal, path.

Our results show that, despite many important differences, quantum
repeater networks behave similarly to classical networks in useful
ways, but the classical principles cannot be applied without thought.
Via detailed physical simulation of both the physics and the classical
messaging protocols, we have investigated several variants and
explored the range of conditions under which these principles apply.

We can predict the \emph{throughput} of a connection based primarily
on the bottleneck link in the path, while the \emph{total work}, in
number of operations performed (pulses or measurements), increases
with the addition of other, non-bottleneck links, much as in a
classical network path.  Applying a form of Dijkstra's algorithm with
the inverse throughput of each hop as the link cost
(Dijkstra/BellGenT) results in strong correlation between our
easily-calculated path cost and actual throughput, and between cost
and total work.  This is achieved with reasonable computational
effort, allowing us to recommend the use of Dijkstra.

In future work, we expect to confirm the general behavior of Dijkstra
with other physical-layer repeater types, adding
single-photon~\cite{PhysRevA.76.012301} and low-photon
number~\cite{munro:PhysRevLett.101.040502} to the qubus systems
explored here.  Our simulator is capable of modeling the important
factor of finite quantum memory lifetimes~\cite{hartmann06}, but the
sheer additional combinatoric complexity that would have come from
including variation of this parameter in both the simulations and
algorithmic arguments in this paper prevents us from presenting these
results.

Perhaps the most important open question is whether these results
apply to error-corrected, rather than purified, repeater
networks~\cite{PhysRevLett.104.180503,PhysRevA.79.032325,munro2010quantum}.
Demonstrating the applicability of Dijkstra to both purify-and-swap
and error-corrected repeater networks would be a strong indicator of
the universality of our results.  We expect that initial network
demonstrations will be purify-and-swap, moving toward error-corrected
but perhaps with a long period of coexistence.

The results presented here, concerning routing within a network of
uniform technology but disparate operating conditions, are part of a
larger program attempting to unify various quantum networking
approaches within a single framework, covering inter-networking
between heterogeneous physical technologies as well as radically
different error correction approaches, routing, multiplexing
approaches and quality of service, and distributed applications.  With
these results, we expect to take advantage of the knowledge
accumulated in half a century of classical networking research and
operation to create a strong quantum network architecture, guide the
experimental focus for the development of repeaters, implement
real-world quantum protocols and accelerate deployment of quantum
networks.

\paragraph{Acknowledgements\\}

This work was supported by the Japan Society for the Promotion of
Science, KAKENHI 21500020.  TDL was partially supported by the
Ministry of Education, Culture, Sports, Science and Technology-Japan
and the National Institute of Information and Communications
Technology.  WJM and KN acknowledge support from the National
Institute of Information and Communications Technology.


\begin{thebibliography}{10}
\bibitem{dijkstra1959ntp}
EW~Dijkstra.
\newblock ``A note on two problems in connexion with graphs,''
\newblock {\em Numerische Mathematik}, 1(1):269--271, 1959.

\bibitem{Chin-WenChou06012007}
Chin-Wen Chou, Julien Laurat, Hui Deng, Kyung~Soo Choi, Hugues de~Riedmatten,
  Daniel Felinto, and H.~Jeff Kimble.
\newblock ``Functional quantum nodes for entanglement distribution over scalable
  quantum networks,''
\newblock {\em Science}, 316(5829):1316--1320, 2007.

\bibitem{kimble08:_quant_internet}
H.~J. Kimble.
\newblock ``The quantum {Internet},''
\newblock {\em Nature}, 453:1023--1030, June 2008.

\bibitem{reichle2006ept}
R.~Reichle, D.~Leibfried, E.~Knill, J.~Britton, RB~Blakestad, JD~Jost,
  C.~Langer, R.~Ozeri, S.~Seidelin, and DJ~Wineland.
\newblock {``Experimental purification of two-atom entanglement,''}
\newblock {\em Nature}, 443(7113):838--41, 2006.

\bibitem{tashima:PhysRevLett.105.210503}
Toshiyuki Tashima, Tsuyoshi Kitano, \ifmmode \mbox{\c{S}}\else
  \c{S}\fi{}ahin~Kaya \"Ozdemir, Takashi Yamamoto, Masato Koashi, and Nobuyuki
  Imoto.
\newblock ``Demonstration of local expansion toward large-scale entangled webs,''
\newblock {\em Phys. Rev. Lett.}, 105(21):210503, Nov 2010.

\bibitem{zhao2003ere}
Z.~Zhao, T.~Yang, Y.A. Chen, A.N. Zhang, and J.W. Pan.
\newblock ``Experimental realization of entanglement concentration and a quantum
  repeater,''
\newblock {\em Physical Review Letters}, 90(20):207901, 2003.

\bibitem{bennett:teleportation}
C.~H. Bennett, G.~Brassard, C.~Cr\'{e}peau, R.~Josza, A.~Peres, and
  W.~Wootters.
\newblock ``Teleporting an unknown quantum state via dual classical and {EPR}
  channels,''
\newblock {\em Physical Review Letters}, 70:1895--1899, 1993.

\bibitem{briegel98:_quant_repeater}
H.-J. Briegel, W.~D\"ur, J.I. Cirac, and P.~Zoller.
\newblock ``Quantum repeaters: the role of imperfect local operations in quantum
  communication,''
\newblock {\em Physical Review Letters}, 81:5932--5935, 1998.

\bibitem{lloyd2004iqi}
S.~Lloyd, J.H. Shapiro, F.N.C. Wong, P.~Kumar, S.M. Shahriar, and H.P. Yuen.
\newblock ``Infrastructure for the quantum {Internet},''
\newblock {\em ACM SIGCOMM Computer Communication Review}, 34(5):9--20, 2004.

\bibitem{bennett:bb84}
C.~H. Bennett and G.~Brassard.
\newblock ``Quantum cryptography: Public key distribution and coin tossing,''
\newblock In {\em Proc. IEEE International Conference on Computers, Systems,
  and Signal Processing}, pages 175--179. IEEE, December 1984.

\bibitem{dodson2009updatingQKD}
D.~Dodson, M.~Fujiwara, P.~Grangier, M.~Hayashi, K.~Imafuku, K.~Kitayama,
  P.~Kumar, C.~Kurtsiefer, G.~Lenhart, N.~Luetkenhaus, et~al.
\newblock {``Updating Quantum Cryptography Report ver. 1,''}
\newblock {\em Arxiv preprint arXiv:0905.4325}, 2009.

\bibitem{lo:qkd-review}
Hoi-Kwong Lo and Yi~Zhao.
\newblock ``Quantum cryptography,''
\newblock In {\em Encyclopedia of Complexity and System Science}. Springer,
  2008.
\newblock arXiv:0803.2507v4 [quant-ph].

\bibitem{elliott:qkd-net}
Chip Elliott, David Pearson, and Gregory Troxel.
\newblock ``Quantum cryptography in practice,''
\newblock In {\em Proc. SIGCOMM 2003}. ACM, ACM, August 2003.

\bibitem{mink09:_qkd_and_ipsec}
Alan Mink, Sheila Frankel, and Ray Perlner.
\newblock ``Quantum key distribution ({QKD}) and commodity security protocols:
  Introduction and integration,''
\newblock {\em International Journal of Network Security \& Its Applications
  (IJNSA)}, 1(2), July 2009.

\bibitem{nagayama09:_ike_for_ipsec_with_qkd}
Shota Nagayama and Rodney Van{ }Meter.
\newblock ``{IKE} for {IPsec} with {QKD},''
\newblock Internet Draft, draft-nagayama-ipsecme-ipsec-with-qkd-00; October 2009, expired
  April 22, 2010.

\bibitem{peev:secoqc}
M~Peev et~al.
\newblock ``The {SECOQC} quantum key distribution network in {Vienna},''
\newblock {\em New Journal of Physics}, 11(7):075001 (37pp), 2009.

\bibitem{chen2010metropolitan}
Teng-Yun Chen, Jian Wang, Hao Liang, Wei-Yue Liu, Yang Liu, Xiao Jiang, Yuan
  Wang, Xu~Wan, Wei-Qi Cai, Lei Ju, Luo-Kan Chen, Liu-Jun Wang, Yuan Gao, Kai
  Chen, Cheng-Zhi Peng, Zeng-Bing Chen, and Jian-Wei Pan.
\newblock {``Metropolitan all-pass and inter-city quantum communication network,''}
\newblock August 2010.
\newblock arXiv:1008.1508v2 [quant-ph].

\bibitem{alleaume:njp-qkd}
R~All\'eaume, F~Roueff, E~Diamanti, and N~L\"utkenhaus.
\newblock ``Topological optimization of quantum key distribution networks,''
\newblock {\em New Journal of Physics}, 11(7):075002, 2009.

\bibitem{ekert1991qcb}
A.K. Ekert.
\newblock ``Quantum cryptography based on {Bell's} theorem,''
\newblock {\em Physical Review Letters}, 67(6):661--663, 1991.

\bibitem{ben-or2005fast}
M.~Ben-Or and A.~Hassidim.
\newblock {``Fast quantum Byzantine agreement,''}
\newblock In {\em Proceedings of the thirty-seventh annual ACM symposium on
  Theory of computing}, pages 481--485. ACM, 2005.

\bibitem{buhrman03:_dist_qc}
Harry Buhrman and Hein R\"ohrig.
\newblock {\em Mathematical Foundations of Computer Science 2003}, chapter
  ``Distributed Quantum Computing,'' pages 1--20.
\newblock Springer-Verlag, 2003.

\bibitem{dhondt05:_dist-qc}
Ellie D'Hondt.
\newblock {\em Distributed quantum computation: A measurement-based approach}.
\newblock PhD thesis, Vrije Universiteit Brussel, July 2005.

\bibitem{tani05:_quant_leader_elect}
Seiichiro Tani, Hirotada Kobayashi, and Keiji Matsumoto.
\newblock ``Exact quantum algorithms for the leader election problem,''
\newblock In {\em Proc. {STACS} 2005: 22nd Annual Symposium on Theoretical
  Aspects of Computer Science}, volume 3404 of {\em Lecture Notes in Computer
  Science}, pages 581--592. Springer-Verlag, 2005.

\bibitem{dur2007epa}
W.~D\"ur and H.J. Briegel.
\newblock {``Entanglement purification and quantum error correction,''}
\newblock {\em Rep. Prog. Phys.}, 70:1381--1424, 2007.

\bibitem{PhysRevLett.104.180503}
Austin~G. Fowler, David~S. Wang, Charles~D. Hill, Thaddeus~D. Ladd, Rodney Van{
  }Meter, and Lloyd C.~L. Hollenberg.
\newblock ``Surface code quantum communication,''
\newblock {\em Phys. Rev. Lett.}, 104(18):180503, May 2010.

\bibitem{PhysRevA.79.032325}
Liang Jiang, J.~M. Taylor, Kae Nemoto, W.~J. Munro, Rodney Van{ }Meter, and
  M.~D. Lukin.
\newblock ``Quantum repeater with encoding,''
\newblock {\em Phys. Rev. A}, 79(3):032325, Mar 2009.

\bibitem{munro2010quantum}
WJ~Munro, KA~Harrison, AM~Stephens, SJ~Devitt, and K.~Nemoto.
\newblock {``From quantum multiplexing to high-performance quantum networking,''}
\newblock {\em Nature Photonics}, 4:792--796, 2010.

\bibitem{wootters:no-cloning}
W.~K. Wootters and W.~H. Zurek.
\newblock ``A single quantum cannot be cloned,''
\newblock {\em Nature}, 299:802, October 1982.

\bibitem{fedrizzi2009high}
A.~Fedrizzi, R.~Ursin, T.~Herbst, M.~Nespoli, R.~Prevedel, T.~Scheidl,
  F.~Tiefenbacher, T.~Jennewein, and A.~Zeilinger.
\newblock {``High-fidelity transmission of entanglement over a high-loss
  free-space channel,''}
\newblock {\em Nature Physics}, 5(6):389--392, 2009.

\bibitem{scheidl2009feasibility}
T.~Scheidl, R.~Ursin, A.~Fedrizzi, S.~Ramelow, X.S. Ma, T.~Herbst, R.~Prevedel,
  L.~Ratschbacher, J.~Kofler, T.~Jennewein, et~al.
\newblock {``Feasibility of 300 km quantum key distribution with entangled
  states,''}
\newblock {\em New Journal of Physics}, 11(085002):085002, 2009.

\bibitem{villoresi08:_space-quantum}
Paolo Villoresi, Thomas Jennewein, Fabrizio Tamburini, Markus Aspelmeyer,
  Cristian Bonato, Rupert Ursin, Claudio Pernechele, Vincenza Luceri, Giuseppe
  Bianco, Anton Zeilinger, and Cesare Barbieri.
\newblock ``Experimental verification of the feasibility of a quantum channel
  between {Space} and {Earth},''
\newblock {\em New Journal of Physics}, 10:033038, 2008.

\bibitem{van-meter07:banded-repeater-ton}
Rodney Van{ }Meter, Thaddeus~D. Ladd, W.~J. Munro, and Kae Nemoto.
\newblock ``System design for a long-line quantum repeater,''
\newblock {\em {IEEE/ACM} Transactions on Networking}, 17(3):1002--1013, June
  2009.

\bibitem{touch2006recursive}
J.D. Touch, Y.S. Wang, and V.~Pingali.
\newblock {``A recursive network architecture,''}
\newblock Technical report, ISI, October 2006.

\bibitem{copsey:q-com-cost}
Dean Copsey, Mark Oskin, Tzvetan Metodiev, Frederic~T. Chong, Isaac Chuang, and
  John Kubiatowicz.
\newblock ``The effect of communication costs in solid-state quantum computing
  architectures,''
\newblock In {\em Proceedings of the fifteenth annual ACM Symposium on Parallel
  Algorithms and Architectures}, pages 65--74, 2003.

\bibitem{isailovic06:_interconnect}
Nemanja Isailovic, Yatish Patel, Mark Whitney, and John Kubiatowicz.
\newblock ``Interconnection networks for scalable quantum computers,''
\newblock In {\em Computer Architecture News, Proc. 33rd Annual International
  Symposium on Computer Architecture}. ACM, June 2006.

\bibitem{1330522}
Tzvetan~S. Metodi, Darshan~D. Thaker, Andrew~W. Cross, Isaac~L. Chuang, and
  Frederic~T. Chong.
\newblock ``High-level interconnect model for the quantum logic array
  architecture,''
\newblock {\em J. Emerg. Technol. Comput. Syst.}, 4(1):1--28, 2008.

\bibitem{oskin:quantum-wires}
Mark Oskin, Frederic~T. Chong, Isaac~L. Chuang, and John Kubiatowicz.
\newblock ``Building quantum wires: The long and short of it,''
\newblock In {\em Computer Architecture News, Proc. 30th Annual International
  Symposium on Computer Architecture}. ACM, June 2003.

\bibitem{van-meter10:dist_arch_ijqi}
Rodney Van{ }Meter, Thaddeus~D. Ladd, Austin~G. Fowler, and Yoshihisa Yamamoto.
\newblock ``Distributed quantum computation architecture using semiconductor
  nanophotonics,''
\newblock {\em International Journal of Quantum Information}, 8:295--323, 2010.

\bibitem{medina2002traffic}
A.~Medina, N.~Taft, K.~Salamatian, S.~Bhattacharyya, and C.~Diot.
\newblock {``Traffic matrix estimation: Existing techniques and new directions,''}
\newblock {\em ACM SIGCOMM Computer Communication Review}, 32(4):174, 2002.

\bibitem{ladd06:_hybrid_cqed}
T.~D. Ladd, P.~van Loock, K.~Nemoto, W.~J. Munro, and Y.~Yamamoto.
\newblock ``Hybrid quantum repeater based on dispersive {CQED} interaction
  between matter qubits and bright coherent light,''
\newblock {\em New Journal of Physics}, 8:184, 2006.

\bibitem{van-loock06:_hybrid_quant_repeater}
P.~van Loock, T.~D. Ladd, K.~Sanaka, F.~Yamaguchi, Kae Nemoto, W.~J. Munro, and
  Y.~Yamamoto.
\newblock ``Hybrid quantum repeater using bright coherent light,''
\newblock {\em Physical Review Letters}, 96:240501, 2006.

\bibitem{childress2006ftq}
L.~Childress, J.M. Taylor, A.S. S{\o}rensen, and M.D. Lukin.
\newblock ``Fault-tolerant quantum communication based on solid-state photon
  emitters,''
\newblock {\em Physical Review Letters}, 96(7):70504, 2006.

\bibitem{munro:PhysRevLett.101.040502}
W.~J. Munro, R.~Van~Meter, Sebastien G.~R. Louis, and Kae Nemoto.
\newblock ``High-bandwidth hybrid quantum repeater,''
\newblock {\em Phys. Rev. Lett.}, 101(4):040502, Jul 2008.

\bibitem{dehaene2003lpp}
J.~Dehaene, M.~Van~den Nest, B.~De~Moor, and F.~Verstraete.
\newblock {``Local permutations of products of Bell states and entanglement
  distillation,''}
\newblock {\em Physical Review A}, 67(2):22310, 2003.

\bibitem{PhysRevA.76.012301}
L.~Jiang, J.~M. Taylor, and M.~D. Lukin.
\newblock ``Fast and robust approach to long-distance quantum communication with
  atomic ensembles,''
\newblock {\em Phys. Rev. A}, 76:012301, Jul 2007.

\bibitem{hartmann06}
L.~Hartmann, B.~Kraus, H.-J. Briegel, and W.~D\"ur.
\newblock ``On the role of memory errors in quantum repeaters,''
\newblock {\em Physical Review A}, 75:032310, 2007.

\end{thebibliography}
%
%

\end{document}